\documentclass[twocolumn,showpacs,preprintnumbers,amsmath,amssymb,prl]{revtex4}

\usepackage{graphicx}
\usepackage{bm}
\usepackage{dcolumn}
\usepackage{amsmath,amsthm,amssymb}
\usepackage{array}

\begin{document}

\title{Ultracold Molecules in the Ro-Vibrational Triplet Ground State}

\author{F. Lang $^{1}$}
\author{K. Winkler $^{1}$}
\author{C. Strauss $^{1}$}
\author{R. Grimm $^{1,2}$}
\author{J. Hecker Denschlag  $^{1}$}
\affiliation{%
$^{1}$Institut f\"ur Experimentalphysik und Zentrum f\"ur Quantenphysik,\\
Universit\"at Innsbruck, A-6020 Innsbruck, Austria\\
$^{2}$Institut f\"ur Quantenoptik und Quanteninformation\\
 der \"Osterreichischen Akademie der Wissenschaften, A-6020 Innsbruck, Austria}

\date{\today}

\begin{abstract}

We report here on the  production of an ultracold gas of tightly
bound Rb$_2$ molecules in the ro-vibrational triplet ground state,
close to quantum degeneracy. This is achieved by optically
transferring weakly bound Rb$_2$ molecules to the absolute lowest
level of the ground triplet potential with a transfer
efficiency of about 90\%. The transfer takes place in a 3D optical
lattice which traps a sizeable fraction of the tightly bound
molecules with a lifetime exceeding 200\,ms.

\end{abstract}
\maketitle

The successful production of quantum degenerate gases of {\em weakly
bound} molecules has triggered a quest for quantum gases of {\em
tightly bound} molecules. These can be used to investigate ultracold
collisions and chemistry of molecules, to produce molecular
Bose-Einstein condensates (BEC), and to develop molecular quantum
optics. Standard laser cooling techniques as developed for atoms~\cite{Met99} do not work for molecules due to their complex internal
structure. Other pathways to cold and dense samples of molecules are
required, such as Stark or Zeeman deceleration~\cite{Mee08,Eli03} and
sympathetic cooling~\cite{Doy04} or association of ultracold atoms
\cite{Jon06,Koh06,Hut06}. Association via Feshbach resonances
\cite{Koh06,Hut06} has directly produced quantum degenerate or near
degenerate ultracold molecular gases
\cite{Her03,Xu03,Joc03,Win05,Vol06}, but
only in very weakly bound states with a high vibrational quantum
number. Furthermore, such molecules are in general unstable when
colliding with each other, particularly if they are composed of
bosonic atoms.

Recently optical schemes have been developed with the goal to
selectively produce cold and dense samples of deeply bound
molecules~\cite{Mil04, Win07, Osp08, Dan08, Vit08, Dei08}, ultimately in a
ro-vibrational ground state. We report here the realization of this
goal by optically transferring a dense ensemble of $^{87}$Rb$_2$
Feshbach molecules to a single quantum level in the ro-vibrational
ground state of the Rb$_2$ triplet potential ($\textrm{a} \ ^3\Sigma_\textrm{u}^+$).
The transfer is carried out in a single step using stimulated Raman adiabatic passage (STIRAP)~\cite{Ber98, Win07, Osp08, Dan08} with an efficiency of almost 90\%, which is only technically limited. The molecules are held in a 3D
optical lattice in which they exhibit a trap lifetime exceeding
200\,ms, after an initial relaxation within 50\,ms.


In contrast to singlet molecules, triplet molecules exhibit a
magnetic moment giving rise to a rich energy level structure in the
presence of magnetic fields. Thus collisions of triplet molecules
should exhibit magnetically tunable scattering resonances, e.g.,
Feshbach resonances. Molecules in the triplet ro-vibrational ground
state can potentially relax to the singlet state $\textrm{X} \ ^1\Sigma_\textrm{g}^+$
through inelastic collisions. This process has not yet been
investigated and can possibly be suppressed. Such a regime would
open interesting prospects for future experiments with molecular
Bose-Einstein condensates and ultracold chemistry~\cite{Kre08}.

\begin{figure}[b]
\centering
\includegraphics[width = 8.5cm]{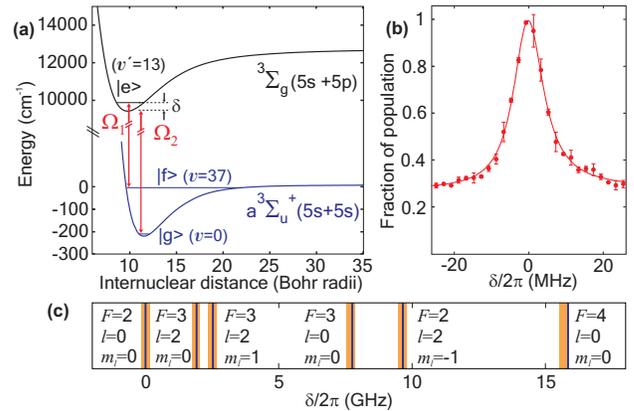}
\caption{(color online) (a) Molecular levels of Rb$_2$. The lasers 1
and 2 couple the molecule levels $|f\rangle$, $|g\rangle$ to the
excited level $|e\rangle$ with Rabi frequencies $\Omega_{1,2} $,
respectively. Note the different energy scales for the ground and excited triplet potentials. (b) Dark resonance. The
data shows the remaining fraction of Feshbach molecules $|f\rangle$
after exposing them to both Raman lasers in a 3\,$\mu$s square pulse.
The two-photon detuning $\delta$ is scanned by varying the
wavelength of laser 2 while keeping laser 1 on resonance. The Rabi
frequencies are $\Omega_{1} = 2\pi\times0.7$\,MHz and $\Omega_{2} =
2\pi\times10$\,MHz. The solid line is a fit from a simple three
level model~\cite{Win07}. (c) Hyperfine and rotational spectrum for
$v$=0 of the $\textrm{a} \ ^3\Sigma_\textrm{u}^+$ potential. The shaded bars correspond to measurements and the widths represent the typical error margin. The thin solid lines are from theoretical calculations and are shown with respective quantum numbers $F,  l$, and its projection $m_l$. The line at $\delta=0$ corresponds to state $|g\rangle$ and is the absolute lowest level of the $\textrm{a} \ ^3\Sigma_\textrm{u}^+$ potential.}
\label{fig:darkstate}
\end{figure}

The starting point for our transfer experiments is a 50$\mu$m-size
pure ensemble of $3\times10^4$ weakly bound Rb$_2$ Feshbach
molecules, produced from an atomic $^{87}$Rb
Bose-Einstein condensate using a Feshbach resonance at a magnetic
field of 1007.4\,G 
(1\,G = 10$^{-4}$\,T). They are
trapped in the lowest Bloch band of a cubic 3D optical lattice with
no more than a single molecule per lattice site~\cite{Tha06} and an
effective lattice filling factor of about 0.3.
The lattice depth for the Feshbach molecules is
60\,$E_r$, where $E_r=\pi^2\hbar^2/2m a^2$ is the recoil energy,
with $m$ the mass of the molecules and $a = 415.22\,$nm the lattice
period. Such deep lattices suppress tunneling between different
sites. The magnetic field is set to 1005.8\,G where the Feshbach
molecules are in a quantum state $|f\rangle$ which correlates with
$|F=2, m_{F}=2, f_{1}=2, f_{2}=2, v=36, l=0\rangle$ at
0\,G~\cite{Lan08}. Here, $F$ and $f_{1,2}$ are the total angular
momentum quantum numbers for the molecule and its atomic
constituents, respectively, and $m_{F}$ is the total magnetic
quantum number; $v$ is the vibrational quantum number for the triplet
potential and $l$ is the quantum number for rotation.

For the transfer we use a stimulated optical Raman transition. Two
lasers (1 and 2) connect the Feshbach molecule level, $|f\rangle$,
via an excited level, $|e\rangle$, to the absolute lowest level in
the triplet potential, $|g\rangle$ (see Fig.\,\ref{fig:darkstate}
(a)). State $|g\rangle$ has a binding energy of
7.0383(2)\,THz$\times h$ and can be described by the quantum numbers
$|F=2, m_{F}=2, S=1, I=3, v=0, l=0\rangle$ where $S$ and
$I$ are the total electronic and nuclear spins of the molecule,
respectively. At 1005.8\,G the ground state is separated by hundreds
of MHz from any other bound level, so there is no ambiguity in what
level is addressed. The level $|e\rangle$ is located in the
vibrational $v$'=13 manifold of the electronically excited
$^3\Sigma_\textrm{g}^+$ (5s + 5p) potential and has $1_\textrm{g}$ character.
It has an excitation energy of 294.6264(2)\,THz$\times h$ with
respect to $|f\rangle$, a width $\Gamma$ = $2\pi \times$8\,MHz and
a Zeeman shift of 3.4\,MHz$\times h$/G. From resonant excitation
measurements we deduce a coupling strength for the transition from
$|f\rangle$ to $|e\rangle$ of
$\Omega_{1}/\sqrt{I_{1}}=2\pi\times$0.4\,MHz/$\sqrt{\mbox{Wcm}^{-2}}$
where $\Omega_{1}$ is the Rabi frequency and $I_{1}$ is the
intensity of laser 1. In comparison, the coupling strength for the
transition from $|g\rangle$ to $|e\rangle$ is
$\Omega_{2}/\sqrt{I_{2}}=2\pi\times$30\,MHz/$\sqrt{\mbox{Wcm}^{-2}}$.
As in Autler-Townes splitting~\cite{Win05}, we deduce
$\Omega_{2}$ from the measured width of a dark resonance which
appears when both lasers resonantly couple to level $|e\rangle$ (see
Fig.~\ref{fig:darkstate}(b)).

\begin{figure}
\includegraphics[width = 8.5cm]{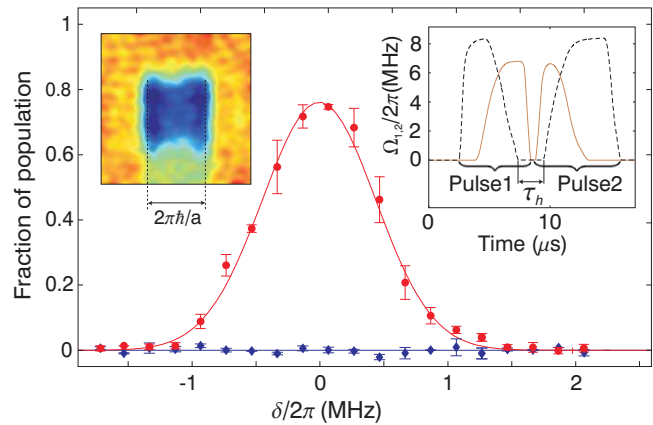}
\centering
\caption{(color online) STIRAP. We plot the efficiency for population transfer from state $|f\rangle$ to state $|g\rangle$ and back with two STIRAP pulses (circles)
as a function of two photon detuning $\delta$. In the dead time between the two STIRAP pulses, no Feshbach molecules can be detected (diamonds). The
continuous lines are from model calculations as described in the text. The
right inset shows the corresponding pulse sequence indicating the Rabi
frequencies of laser 1 (solid line) and laser 2 (dashed line). The
hold time $\tau_h$, the time between the actual population transfers, is equal
to $2\,\mu$s. The left inset is a absorption image which displays
the atomic quasi-momentum distribution in the optical lattice after the
round-trip STIRAP transfer and subsequent adiabatic molecule dissociation.
Atoms located in the inner square stem from the lowest Bloch band.
$2\pi\hbar/a$ is the modulus of the reciprocal lattice vector.}
\label{fig:transfer}
\end{figure}

The positions of the deeply bound energy levels of the Rb$_2$
triplet potentials $^3\Sigma_\textrm{g}^+$ and $\textrm{a} \ ^3\Sigma^{+}_\textrm{u}$ were not
precisely known before this work. Therefore, we have carried out
extensive single- and two-color spectroscopy on our pure ensemble of
Feshbach molecules. We have mapped out the vibrational progression
of both potentials to the ground state $v$ ($v$') = 0  and find good
agreement with theoretical calculations based on {\em ab-initio}
potentials~\cite{Par01}. In order to determine the hyperfine and rotational
structure of the $\textrm{a} \ ^3\Sigma^{+}_\textrm{u}$ vibrational ground state, we use a different intermediate level in the $^3\Sigma_\textrm{g}^+$ potential instead of $|e \rangle$. It has $0_\textrm{g}^-$ character and quantum number $I = 3$. Due to the selection rule $\Delta I = 0$ and the fact that $I$ is a good quantum number for the deeply bound $\textrm{a} \ ^3\Sigma^{+}_\textrm{u}$ states, this considerably restricts the number of observed lines. We find excellent agreement of the measured data with theoretical calculations based on a close-coupled channel model with essentially no free parameters (see
Fig.~\ref{fig:darkstate}(c)). In particular we identify the lowest
observed state $|g\rangle$ as absolute ground state of the $\textrm{a} \
^3\Sigma^{+}_\textrm{u}$ potential. A detailed discussion of further
spectroscopic measurements as well as their analysis will be
presented elsewhere.


STIRAP is a very efficient transfer method based on a stimulated
Raman transition. It uses a counter-intuitive pulse sequence during
which molecules are kept in a dynamically changing dark
superposition state $|\Psi_{ds}\rangle = \left({\Omega_2|f\rangle -
\Omega_1|g\rangle}\right)/{\sqrt{\Omega_1^2+\Omega_2^2}}$. This dark
state is decoupled from the light in the sense that there is no
excitation of the short lived state $| e\rangle$, which suppresses
losses during transfer (see e.\,g., the dark resonance in
Fig.~\ref{fig:darkstate}(b)). A vital condition for STIRAP is the
relative phase stability between the two Raman lasers. Both of our
Raman lasers, a Ti:Sapphire laser at 1017.53~nm (laser 1) and a grating-stabilized diode laser at 993.79~nm (laser 2) are Pound-Drever-Hall locked to a single cavity which itself is locked to an atomic $^{87}$Rb-line. From the
lock error signals, we estimate frequency stabilities on a
ms-timescale of 40\,kHz and 80\,kHz for lasers 1 and 2,
respectively. Thus, the transfer has to take place on a $\mu$s
timescale in order not to lose phase coherence during STIRAP. Both
laser beams have a waist of 130\,$\mu$m at the location of the
molecular sample, propagate collinearly, and are polarized parallel
to the direction of the magnetic bias field. Thus the lasers can
only induce $\pi$ transitions.

We perform STIRAP by adiabatically ramping the Raman laser intensities as shown in the
right inset of Fig.~\ref{fig:transfer}.
Pulse 1 efficiently transfers the molecules from $|f\rangle$ to $|g\rangle$. In order to
detect the molecules in state $|g\rangle$ after the transfer, we bring them back to
$|f\rangle$ with a second, reversed STIRAP pulse sequence. We then dissociate the
molecules into pairs of atoms at the Feshbach resonance. By releasing these atoms from
the optical lattice in the manner described in~\cite{Win06}, we can map out the Bloch
bands in momentum space. After 13\,ms of ballistic expansion the corresponding atomic
distribution is recorded with standard absorption imaging (see left inset in
Fig.~\ref{fig:transfer}). For our signals, we only count atoms in the central square
zone, corresponding to the lowest Bloch band~\cite{Win06}.

Figure~\ref{fig:transfer} shows the total transfer efficiency after
two STIRAP transfers which are separated by a hold time $\tau_h =
2\,\mu$s. The transfer efficiency for this {\em round trip} STIRAP
process is plotted as a function of the two-photon detuning $\delta$
and reaches about 75\% at resonance ($\delta = 0$). Assuming equal
efficiencies for both transfers, this corresponds to a single
transfer efficiency of 87\% and a total number of $2.6\times10^4$
molecules in state $|g\rangle$. We have experimentally verified that
no molecules remain in state $|f\rangle$ between the two STIRAP
pulses (diamonds in Fig.~\ref{fig:transfer}). Any such
molecules would quickly be removed by laser 1 at the end of the
first STIRAP pulse, which is kept on at maximum power for 1$\mu$s
after ramping down laser 2. Thus, all molecules that are retrieved
after the second STIRAP transfer have been deeply bound in state
$|g\rangle$. The 1\,MHz width (FWHM) of the transfer efficiency is
determined by power- and Fourier-broadening~\cite{com01} and is in
good agreement with a 3-level model (see solid lines in
Fig.~\ref{fig:transfer}). We use a master equation which takes into
account decoherence due to phase fluctuations of the Raman lasers
\cite{Wal94}. These fluctuations can be expressed in terms of a
short-term relative linewidth of the lasers, $\gamma$, which from
fits we determine to be about $2\pi\times$20\,kHz. Our calculations
indicate that half of the losses are due to nonadiabaticity and half
are due to the non-ideal laser system. In principle, losses could
also be due to other effects, such as coupling to levels outside of $|f\rangle$, $|e\rangle$ and
$|g\rangle$. However, we have verified that this is not the case,
because losses due to optical excitation are completely negligible
when we expose a pure ensemble of $|f\rangle$ ($|g\rangle$)
molecules only to laser 2 (1). In addition, we did not detect any laser
power dependent shift of the two-photon resonance within the
accuracy of our measurements of $2\pi\times$\,200\,kHz.

\begin{figure}
\includegraphics[width = 8.5cm]{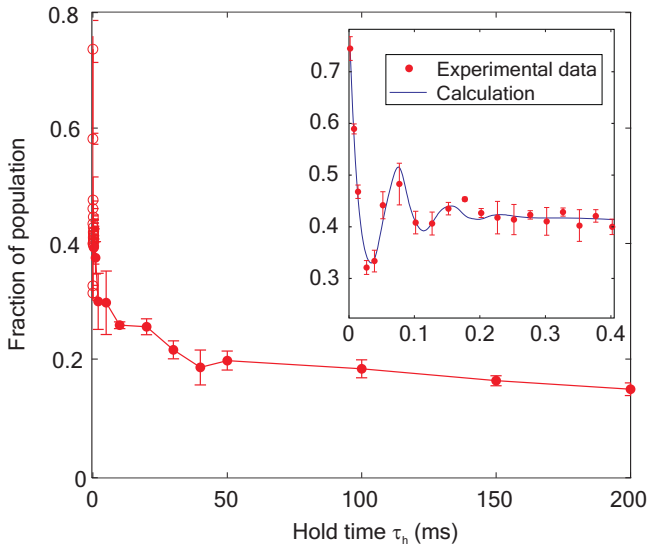}
\centering
\caption{(color online) Dynamics and lifetime of $|g\rangle$  molecules in the optical lattice.
We plot the transfer efficiency for the round trip STIRAP process as a function
of the hold time $\tau_h$ between the two STIRAP pulses. We only count
molecules whose constituent atoms end up in the lowest Bloch band after
transfer (see left inset of Fig.~\ref{fig:transfer}). Except for $\tau_h$, all
other experimental parameters are the same as in Fig.~\ref{fig:transfer}.
Molecules are lost on three different timescales, 100\,$\mu$s, 50\,ms, and
$\approx$200\,ms. The inset zooms into the first 400\,$\mu$s. The oscillations in the transfer efficiency are due to breathing oscillations of localized spatial
wavepackets of molecules in the lattice sites. The solid line is from a
multi-band model calculation. The data shown in the inset are plotted with open
plot symbols in the main plot. The line connects neighboring data points.}
\label{fig:lifetime_gs}
\end{figure}

We also investigate the dynamics and lifetime of the deeply bound molecules in the
optical lattice. Due to their strong binding, molecules in state $|g\rangle$ cannot be
expected to have a polarizability similar to that of Feshbach molecules, and it is not clear {\em a priori} what strength or even sign the optical lattice potential will have for them.
Indeed, as we show below, the lattice potential is attractive for the $|g\rangle$
molecules, but a factor 10$\pm$2 shallower compared to the potential for the Feshbach
molecules. Repeating the transfer experiment, we now vary the hold time $\tau_h$ between
the two STIRAP transfers (see Fig.~\ref{fig:lifetime_gs}). Interestingly, for short hold
times the transfer efficiency exhibits a damped oscillation (see inset). The period and
damping time are both about 80\,$\mu$s. After 250\,$\mu$s the efficiency levels off at
40\% and then decays much more slowly. The initial oscillation can be understood as
follows. We consider the localized spatial wavepacket of a Feshbach molecule at a
particular lattice site in the lowest Bloch band. The first STIRAP transfer projects this
wavepacket onto the much shallower lattice potential felt by the $|g\rangle$ molecules.
As a consequence $|g\rangle$ molecules are coherently spread over various Bloch bands,
and the wavepacket undergoes ``breathing'' oscillations with the lattice site trap
frequency $\omega_t$. These coherent oscillations are damped by tunneling of $|g\rangle$
molecules in higher Bloch bands to neighboring lattice sites. The reverse STIRAP transfer
maps this periodic oscillation back to the Feshbach molecules. Higher Bloch bands are
populated here as well, but are at most partially counted in our scheme, which leads to
an apparent loss of our transfer efficiency.
We can describe the data well using a 3D multi-band model  (see
solid line, inset Fig.~\ref{fig:lifetime_gs}). In this model
the states $|f\rangle$, $|e\rangle$ and $|g\rangle$ have a
substructure due to the Bloch bands of the optical lattice and the
resulting levels are coupled by the laser fields. From fits to the
data we extract the trap frequency $\omega_t$ in a single lattice
site, which determines the lattice depth for the molecules in state
$|g\rangle$.
We note that the earlier analysis of Fig.~\ref{fig:transfer} does not include optical
lattice effects. However, because the hold time $\tau_h$ is so short (2 $\mu$s), molecule
signal losses due to oscillation amount to only 4\%. In fact, the multi-band model leads to the same theoretical curve as
shown when we use a short-term relative laser line width $\gamma  = 2\pi \times 18 $ kHz, close to the previous value.

For longer hold times of up to 200\,ms, Figure~\ref{fig:lifetime_gs} shows the
time dependent loss of the deeply bound molecules. Within the first 50\,ms the
fraction of recovered molecules drops to 20\%. We attribute this loss mainly to
the fact that all molecules in excited bands will simply fall out of the
lattice since they are essentially unbound. For the remaining molecules in the
lowest band we find a lifetime exceeding our maximum experimental observation
time which is limited due to heating of the magnetic field coils.

To conclude, using a nearly 90\% efficient STIRAP transfer we have created a dense and ultracold ensemble of deeply bound Rb$_2$ molecules in the absolute lowest
quantum state of the $\textrm{a} \ ^3\Sigma_\textrm{u}^+$ potential. These deeply bound molecules
were trapped in a 3D optical lattice and we observed coherent motional dynamics
of their spatial wavepackets in the sites. This indicates that besides the
internal degrees of freedom,  the external degrees of freedom are also
precisely defined after transfer. The transfer of molecules into a single Bloch
band should be possible, either by matching the lattice depths of weakly and
deeply bound molecules, or by spectroscopically resolving the Bloch
bands~\cite{Rom04}. The latter involves longer STIRAP pulses and more tightly
phase-locked Raman lasers, with the added benefit of increasing the transfer
efficiency further. Investigating the collisional behavior of the triplet
molecules will be the next goal as it is of central importance for ultracold
chemistry~\cite{Kre08} and for achieving molecular BEC. An appealing
 way to reach BEC is by melting an optical-lattice-induced
Mott insulator of ro-vibrational ground state molecules
\cite{Jak02}. For this, we have to improve the lattice occupation of
our initial ensemble of Feshbach molecules~\cite{Vol06} and use a
selective STIRAP transfer to the lowest Bloch band.

During the preparation of our manuscript, we learned that ro-vibrational ground state molecules have been produced with KRb~\cite{Ni08}.

\begin{acknowledgments}
The authors thank Birgit Brandst\"atter, Olivier Dulieu, Paul Julienne, Christiane Koch, Roman Krems, Helmut Ritsch, Peter v.d.Straten, and Eberhard Tiemann for theoretical support. We acknowledge the usage of a close-coupled channel code from NIST. We thank Devang Naik, Tetsu Takekoshi and Gregor Thalhammer for help in the lab as well as Florian Schreck and the group of Hanns-Christoph N\"agerl for helpful exchange. This work was supported by the Austrian Science Fund (FWF) within SFB 15 (project part 17).


\end{acknowledgments}

\end{document}